\documentclass[a4paper,11pt]{article}
\usepackage{pos}

\usepackage{physics}

\usepackage[compat=1.1.0]{tikz-feynhand}
\usetikzlibrary{patterns}
\usetikzlibrary{decorations.markings}

\newcommand{\1}{\mbox{1}\hspace{-0.25em}\mbox{l}}

\title{Lattice construction of mixed 't~Hooft anomaly with higher-form symmetry}

\author*[a]{Motokazu Abe}
\author[b]{Okuto Morikawa}
\author[a]{Soma Onoda}
\author[a]{Hiroshi Suzuki}
\author[c]{Yuya~Tanizaki}

\affiliation[a]{Department of Physics, Kyushu University,\\
744 Motooka, Nishi-ku, Fukuoka 819-0395, Japan}

\affiliation[b]{Department of Physics, Osaka University,\\
Toyonaka, Osaka 560-0043, Japan}

\affiliation[c]{Yukawa Institute for Theoretical Physics, Kyoto University\\
Kitashirakawa Oiwakecho, Sakyo-ku, Kyoto 606-8502, Japan}

\emailAdd{abe.motokazu@phys.kyushu-u.ac.jp}
\emailAdd{o-morikawa@het.phys.sci.osaka-u.ac.jp}
\emailAdd{onoda.soma.270@s.kyushu-u.ac.jp}
\emailAdd{hsuzuki@phys.kyushu-u.ac.jp}
\emailAdd{yuya.tanizaki@yukawa.kyoto-u.ac.jp}

\abstract{
In this talk, we give the lattice regularized formulation of the mixed 't~Hooft anomaly between the $\mathbb{Z}_N$ $1$-form symmetry and the $\theta$ periodicity for $4$d pure Yang-Mills theory, which was originally discussed by Gaiotto \textit{et al.}\ in the continuum description. 
For this purpose, we define the topological charge of the lattice $SU(N)$ gauge theory coupled with the background $\mathbb{Z}_N$ $2$-form gauge fields $B_p$ by generalizing L\"uscher's construction of the $SU(N)$ topological charge. 
We show that this lattice topological charge enjoys the fractional $1/N$ shift completely characterized by the background gauge field $B_p$, and this rigorously proves the mixed 't~Hooft anomaly with the finite lattice spacings. 
As a consequence, the Yang-Mills vacua at $\theta$ and $\theta+2\pi$ are distinct as the symmetry-protected topological states when the confinement is assumed. 
}

\FullConference{The 40th International Symposium on Lattice Field Theory (Lattice 2023)\\
July 31st - August 4th, 2023\\
Fermi National Accelerator Laboratory\\}

\begin{document}
\maketitle

\section{Introduction}
Symmetry is one of the most important ideas in physics.
It is also quite essential to understand a phenomenon of symmetry breaking, e.g., the quantum anomaly.
When a gauge symmetry, to which a global symmetry is promoted, is anomalous, such an anomaly is called the 't~Hooft anomaly~\cite{tHooft:1979rat}.
The 't~Hooft anomaly is invariant under the renormalization group and its matching condition between different energy scales plays an important role to restrict the low energy dynamics of gauge theories. 
It is quite interesting to apply this idea to nonperturbative phenomena in strongly coupled field theories such as QCD.

Recently, the notion of symmetry has been largely generalized~\cite{Gaiotto:2014kfa}. In recent studies, the existence of a symmetry is regarded as being equivalent to the existence of a topological defect. Here, the topological defect is an object such that it is invariant under the deformation of a surface to which the definition of the defect refers.
For traditional symmetries, the surface of the defect is co-dimension $1$. As a generalization, we define a $p$-form symmetry on a co-dimension $(p+1)$ surface. Let us focus on the $SU(N)$ gauge theory with the $\theta$ term. This theory has the mixed 't~Hooft anomaly between the $\mathbb{Z}_N$ $1$-form symmetry and the time-reversal symmetry when $\theta=\pi$~\cite{Gaiotto:2017yup}. Then, one finds that the phase structures at $\theta=0$ and $2\pi$ are distinguishable according to monopole or dyon condensation.
Now, the important point is that, in this theory coupled to $\mathbb{Z}_N$ $2$-form gauge fields associated with the $1$-form symmetry, the topological charge becomes fractional instead of integral.\footnote{This fractional topological charge is well-known for long time~\cite{t-Hooft:1980aa,vanBaal:1982ag} and the fractionality is caused by twisted boundary conditions associated with the magnetic flux.}

Various studies of anomaly matching associated with the generalized symmetry are vigorously carried out in many quantum field theories (QFTs). However, since QFTs include an infinite number of degrees of freedom, it is difficult to fully define QFTs with interactions mathematically as it stands. Due to this issue, the perturbation theory, which is a promising technique in a broad sense of physical systems, suffers from ultraviolet divergences in perturbative coefficients. One may use a regularization method based on the perturbation theory while it is too tough to address nonperturbative phenomena, e.g., the quark confinement in QCD. A well-established nonperturbative regularization is now provided by lattice gauge theory.

In this talk, we aim to understand the above consequences through 't~Hooft anomaly matching with generalized symmetries in completely regularized theories. A gauge-invariant nonperturbative regularization is provided by lattice gauge theory. First, we review L\"uscher's construction of the topological sectors on the lattice in Sect.~\ref{sec:Luscher}. Then, in Sect.~\ref{sec:fractional_TC}, extending this L\"uscher's construction, we construct the fractional topological charge in the lattice $SU(N)$ gauge theory coupled with $2$-form gauge fields.\footnote{
    We also construct the fractional topological charge with the $U(1)/\mathbb{Z}_q$ principal bundle on the lattice gauge theory~\cite{10.1093/ptep/ptad009}.
}
In addition, this fractional topological charge is manifestly invariant under the $\mathbb{Z}_N$ $1$-form symmetry. By using this lattice fractional topological charge, we obtain the lattice action with the $\theta$ term and derive the anomalous relation for the Yang--Mills partition function on the lattice~\cite{Abe:2023aa}.

\section{Review of L\"uscher's construction of the lattice topological charge}\label{sec:Luscher}
In the lattice gauge theory, basic degrees of freedom are link variables $U_l=U(n,\mu)$, where $l$ means a link connecting two sites $n$ and $n+\hat{\mu}$. Here, we define the minimal constituents which are invariant under the $SU(N)$ gauge transformation: $U(n,\mu)\mapsto g_n^{-1}U(n,\mu) g_{n+\hat{\mu}}$ for $g_n\in SU(N)$,
\begin{align}
    U_p=U_{\mu\nu}(n)
    \equiv \mathcal{P}\prod_{l\in \partial p}U_l
    =U(n,\mu)U(n+\hat{\mu},\nu)U(n+\hat{\nu},\mu)^{-1}U(n,\nu)^{-1} ,
\end{align}
where $p$ denotes a plaquette. For this plaquette variable $U_p$, the $SU(N)$ gauge transformation acts as $U_p\mapsto g_n^{-1}U_p g_n$ where $n$ denotes the initial and final point of the closed loop for $p$. At this time, the value $\tr(U_p)$ is invariant under the $SU(N)$ gauge transformation. Then, we define the Wilson plaquette action as 
\begin{align}
    S_W[U_l] \equiv \sum_p\beta\left[ \tr\left( \1-U_p \right)+\mathrm{c.c.} \right].
    \label{eq:Wilson_action}
\end{align}
where $\beta$ is the coupling constant for the lattice $SU(N)$ gauge theory. This action is manifestly invariant under the $SU(N)$ gauge transformation.

\begin{figure}[htbp]
      \centering
      \begin{tikzpicture}[scale=0.6]
          \draw (0,0,0)--(0,4,0)--(4,4,0)--(4,0,0)--(0,0,0);
          \draw (4,4,0)--(4,4,-4)--(0,4,-4)--(0,4,0);
          \draw (4,0,0)--(8,0,0)--(8,4,0);
          \draw (4,4,0)--(8,4,0)--(8,4,-4)--(4,4,-4);
          \draw (8,0,0)--(8,0,-4)--(8,4,-4);
          \draw[dashed](0,0,0)--(0,0,-4)--(0,4,-4);
          \draw[dashed](0,0,-4)--(4,0,-4)--(4,0,0);
          \draw[dashed](4,4,-4)--(4,0,-4)--(8,0,-4);
          \draw [red!15!orange, very thick] (3.95,0,0)--(3.95,4,0);
          \draw [->,blue, very thick] (4,0,0.05)--(6,0,0.05);
          \draw [blue, very thick] (6,0,0.05)--(8,0,0.05);
          \draw [red!40!blue, opacity=0.7, arrows = {->[slant=0.5]}, line width = 6] (3,2,-2) -- (5,2,-2);
          \fill[pattern=north east lines, pattern color = red](4,0,0)--(4,0,-4)--(4,4,-4)--(4,4,0);
          \node [red] at (5.2,3.5,-3) {$f(n,\mu)$};
          \node at (4,0,0) [below] {$n$};
          \node [red!15!orange] at (4,2,0) [left] {$p(n,\mu,\nu)$};
          \node [blue] at (6,0,0) [above] {$U(n,\mu)$};
          \node [red!40!blue] at (5,2,-2) [right] {$v_{n,\mu}(x)$};
          \node at  (2.6,4,-3) {$c(n-\Hat{\mu})$};
          \node at (7,4,-3) {$c(n)$};
          \draw[->,>=stealth] (-3,0,0) -- (-1,0,0) node [right]{$x_\mu$};
          \draw[->,>=stealth] (-3,0,0) -- (-3,0,-2) node [above]{$x_\nu$};
          \draw[->,>=stealth] (-3,0,0) -- (-3,2,0) node [above]{$x_\rho$};
      \end{tikzpicture}
      \caption{Illustration for the definition of variables on the lattice such as $c(n)$, $f(n,\mu)$, $p(n,\mu,\nu)$.}
      \label{fig:lattice_patch}
\end{figure}
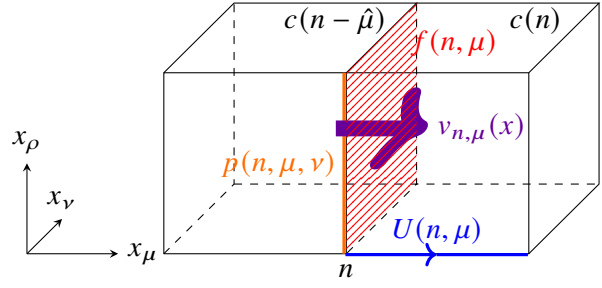

Now, let us add the so-called $\theta$~term, $i\theta Q_{\mathrm{top}}$, to the above Wilson action, where $Q_{\mathrm{top}}$ is the topological charge defined below. Note that this topological term acts only on the boundary, but actually it makes the structure of the original theory quite rich owing to nontrivial homotopy such as instanton. With a set of hyper-cubic unit cells as a special covering, the definition of topological charge is given by~\cite{Luscher:1981zq,vanBaal:1982ag}\footnote{
    In this paper, we work on the $4$-torus $T^4$ and its hyper-cubic lattice discretization $\Lambda_L$. In addition, we define some variables on the hyper-cubic structure as follows: The unit cell is denoted as 
    \begin{align}
        c(n)
        =\left\{x\in T^4\bigm|
        \text{$0\le x_\mu-n_\mu\le1$ for $\mu=1$, \dots, $4$}\right\}
    \end{align}
    for~$n\in\Lambda_L$, and we would like to define the transition function on
    each face
    $f(n,\mu)=c(n)\cap c(n-\Hat{\mu})$, 
    where $\Hat{\mu}$ is the unit vector along the $\mu$th direction. Moreover, we define the cocycle condition on $p(n,\mu,\nu)=c(n)\cap c(n-\Hat{\mu})\cap c(n-\Hat{\mu})\cap c(n-\Hat{\mu}-\Hat{\nu})$ as follows:
    \begin{align}
        v_{n-\hat{\nu},\mu}(n)v_{n,\nu}(n)v_{n,\mu}(n)^{-1}v_{n-\hat{\mu},\nu}(n)^{-1}=\1.
    \end{align}
}
\begin{align}
    Q_{\mathrm{top}}[U_\ell]\equiv\sum_{n\in\Lambda_L}q(n),
 \end{align}
where
\begin{align}
    q(n)
    &=-\frac{1}{24\pi^2}
    \sum_{\mu,\nu,\rho,\sigma}
    \varepsilon_{\mu\nu\rho\sigma}
    \int_{f(n,\mu)}\dd[3]{x}
    \tr\left[
    (v_{n,\mu}^{-1}\partial_\nu v_{n,\mu})
    (v_{n,\mu}^{-1}\partial_\rho v_{n,\mu})
    (v_{n,\mu}^{-1}\partial_\sigma v_{n,\mu})
    \right]
 \notag\\
    &\qquad{}
    -\frac{1}{8\pi^2}
    \sum_{\mu,\nu,\rho,\sigma}\varepsilon_{\mu\nu\rho\sigma}
    \int_{p(n,\mu,\nu)}\dd[2]{x}
    \tr\left[
    (v_{n,\mu}\partial_\rho v_{n,\mu}^{-1})
    (v_{n-\Hat{\mu},\nu}^{-1}\partial_\sigma v_{n-\Hat{\mu},\nu})
    \right].
 \label{eq:TopologicalChargeDensity}
\end{align}
Here, $v_{n,\mu}(x)$ denotes the transition function at $x$ in the overlap $f(n,\mu)$ of patches beside each other as illustrated in Fig.~\ref{fig:lattice_patch}. Naively it is a gauge transformation function from the ``gauge field'' at~$x\in c(n-\Hat{\mu})$ to that at~$x\in c(n)$, or rather we construct $v_{n,\mu}(x)$ at any~$x$ from $U_\ell$. Equation~\eqref{eq:TopologicalChargeDensity} provides the definition of the lattice topological charge and reproduces $\frac{1}{16\pi^2}\int_{T^4} \dd[4]{x}\varepsilon_{\mu\nu\rho\sigma}\tr\left( F_{\mu\nu}F_{\rho\sigma} \right)\in\mathbb{Z}$ in the continuum limit, where $F_{\mu\nu}$ is a field strength.

There are, however, two obstacles to construct $v_{n,\mu}(x)$.
First, a configuration of lattice fields can be connected to other configurations under continuum deformations, and  hence the lattice topological charge is not necessarily an integer. How can we calculate the integral topological charge on the lattice? The answer has already been presented by L\"uscher~\cite{Luscher:1981zq}. Restricting the gauge field on the lattice to being smooth, we can find that the topological charge on the lattice becomes an integer. This condition of smoothness is called the admissibility condition:
\begin{align}
    \left\|\1-U_p\right\|<\varepsilon,
 \label{eq:admissibility}
\end{align}
where $\|\cdot\|$ refers to the matrix norm and $\varepsilon$ is any small positive number independent of the coupling constant and the size of the lattice. For instance, it is known that in the $SU(2)$ gauge theory, $\varepsilon=0.03$.

Secondly, in order to calculate the topological charge, the transition function $v_{n,\mu}$ in Eq.~\eqref{eq:TopologicalChargeDensity} should be the continuum function of the coordinate $x$ not only of the corner $n$. Here, it is straightforward for $v_{n,\mu}(n)$ to be identical to $U(n-\Hat{\mu},\mu)$. Again, L\"uscher provided the explicit expression of~$v_{n,\mu}(x)$ in terms of the interpolation function $S^m_{n,\mu}(x)$ ($m=n$ or $n-\hat{\mu}$) as follows:
\begin{align}
    v_{n,\mu}(x)=S^{n-\hat{\mu}}_{n,\mu}(x)^{-1}v_{n,\mu}(n)S^n_{n,\mu}(x),
    \label{eq:interpolate_v}
\end{align}
where
\begin{align}
    f^m_{n,\mu}(x_\gamma)
        =&(u^m_{s_3s_0})^{y_\gamma}(u^m_{s_0s_3}u^m_{s_3s_7}u^m_{s_7s_2}u^m_{s_2s_0})^{y_\gamma}u^m_{s_0s_2}(u^m_{s_2s_7})^{y_\gamma},\\
    g^m_{n,\mu}(x_\gamma)
        =&(u^m_{s_5s_1})^{y_\gamma}(u^m_{s_1s_5}u^m_{s_5s_4}u^m_{s_4s_6}u^m_{s_6s_1})^{y_\gamma}u^m_{s_1s_6}(u^m_{s_6s_4})^{y_\gamma},\\
    h^m_{n,\mu}(x_\gamma)
        =&(u^m_{s_3s_0})^{y_\gamma}(u^m_{s_0s_3}u^m_{s_3s_5}u^m_{s_5s_1}u^m_{s_1s_0})^{y_\gamma}u^m_{s_0s_1}(u^m_{s_15})^{y_\gamma},\\
    k^m_{n,\mu}(x_\gamma)
        =&(u^m_{s_7s_2})^{y_\gamma}(u^m_{s_2s_7}u^m_{s_7s_4}u^m_{s_4s_6}u^m_{s_6s_2})^{y_\gamma}u^m_{s_2s_6}(u^m_{6s_4})^{y_\gamma},\\
    l^m_{n,\mu}(x_\beta,x_\gamma)
        =&\left[ f^m_{n,\mu}(x_\gamma)^{-1} \right]^{y_\beta}\left[ f^m_{n,\mu}(x_\gamma)k^m_{n,\mu}(x_\gamma)g^m_{n,\mu}(x_\gamma)^{-1}h^m_{n,\mu}(x_\gamma)^{-1} \right]^{y_\beta}\nonumber\\
        &\cdot h^m_{n,\mu}(x_\gamma)\left[ g^m_{n,\mu}(x_\gamma) \right]^{y_\beta},\\
    S^m_{n,\mu}(x_\alpha,x_\beta,x_\gamma)
        =&(u_{s_0s_3}^m)^{y_\gamma}\left[ f^m_{n,\mu}(x_\gamma) \right]^{y_\beta}\left[ l^m_{n,\mu}(x_\beta,x_\gamma) \right]^{y_\alpha}.
        \label{eq:S}
\end{align}
Now, we defer to Sect.~3 in Ref.~\cite{Luscher:1981zq} to look at the detail of the meaning and definition of these functions like $u^m_{s_0s_3}$.  We refer briefly to the meaning of the interpolation function $S^m_{n,\mu}(x)$. This function connects the site $m$ to~$x$, and then, Eq.~\eqref{eq:interpolate_v} means the connection as $x\to  n-\Hat{\mu} \to n \to x$. That is, $v_{n,\mu}(x)$ means the transition function on~$x$.

By using this interpolated transition function, we can check that the cocycle condition for~$v_{n,\mu}(x)$ holds in the same way as the original cocycle condition of the corner~$n$. As a result, L\"uscher~\cite{Luscher:1981zq} proved the cocycle condition at any~$x$,
\begin{align}
    v_{n-\hat{\nu},\mu}(x)v_{n,\nu}(x)v_{n,\mu}(x)^{-1}v_{n-\hat{\mu},\nu}(x)^{-1}=\1.
\end{align}

\section{Fractional topological charge on the lattice}\label{sec:fractional_TC}
In the previous section, we have reviewed the construction of the integral topological charge on the lattice. From now on, in order to calculate the fractional topological charge on the lattice, we extend the above definition of the transition function $v_{n,\mu}(x)$ to $\Tilde{v}_{n,\mu}(x)$ which behaves covariantly under the $\mathbb{Z}_N$ $1$-form gauge transformation.\footnote{
    In what follows, variables with a tilde are subject to the $1$-form gauge covariance, which differ from L\"uscher's construction.
}
\subsection{Covariant transition function and calculation of the topological charge}
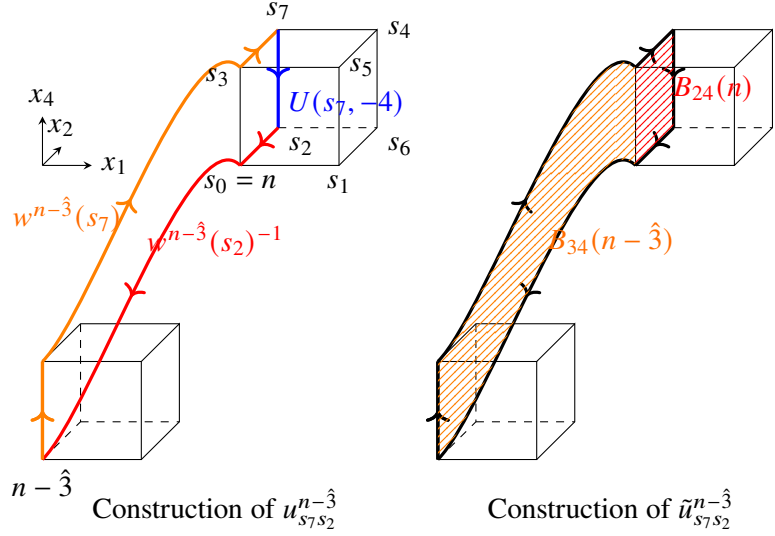
\begin{figure}[ht]
    \centering
    \begin{tikzpicture}[scale=0.65]
        \foreach \x in {0,1}{
            \foreach \y in {0,1}{
                \draw (0-4*\x+8*\y,0-6*\x,0)--(0-4*\x+8*\y,2-6*\x,0)--(2-4*\x+8*\y,2-6*\x,0)--(2-4*\x+8*\y,0-6*\x,0)--(0-4*\x+8*\y,0-6*\x,0);
                \draw (2-4*\x+8*\y,0-6*\x,0)--(2-4*\x+8*\y,0-6*\x,-2)--(2-4*\x+8*\y,2-6*\x,-2)--(2-4*\x+8*\y,2-6*\x,0);
                \draw (2-4*\x+8*\y,2-6*\x,-2)--(0-4*\x+8*\y,2-6*\x,-2)--(0-4*\x+8*\y,2-6*\x,0);
                \draw[dashed](0-4*\x+8*\y,0-6*\x,0)--(0-4*\x+8*\y,0-6*\x,-2)--(0-4*\x+8*\y,2-6*\x,-2);
                \draw[dashed](0-4*\x+8*\y,0-6*\x,-2)--(2-4*\x+8*\y,0-6*\x,-2);
            };
        };
        \begin{scope}[decoration={markings,mark=at position 0.5 with {\arrow{>}}}] 
            \draw [very thick,postaction={decorate},orange] (-4,-6,0)--(-4,-4,0);
            \draw [very thick,postaction={decorate},orange] (-4,-4,0) .. controls (-3,-3,0) and (-1,3,0) .. (0,2,0);
            \draw [very thick,postaction={decorate},red]  (0,0,0) .. controls (-1,1,0) and (-3,-5,0)..(-4,-6,0);
            \draw [very thick,postaction={decorate},orange] (-4,-6,0)--(-4,-4,0);
            \draw [very thick,postaction={decorate},orange] (0,2,0)--(0,2,-2);
            \draw [very thick,postaction={decorate},blue] (0,2,-2)--(0,0,-2);
            \draw [very thick,postaction={decorate},red] (0,0,-2)--(0,0,0);
        \end{scope}
        \begin{scope}[decoration={markings,mark=at position 0.5 with {\arrow{>}}}] 
            \draw [very thick,postaction={decorate}] (4,-6,0)--(4,-4,0);
            \draw [very thick,postaction={decorate}] (4,-4,0) .. controls (5,-3,0) and (7,3,0) .. (8,2,0);
            \draw [very thick,postaction={decorate}]  (8,0,0) .. controls (7,1,0) and (5,-5,0)..(4,-6,0);
            \draw [very thick,postaction={decorate}] (4,-6,0)--(4,-4,0);
            \draw [very thick,postaction={decorate}] (8,2,0)--(8,2,-2);
            \draw [very thick,postaction={decorate}] (8,2,-2)--(8,0,-2);
            \draw [very thick,postaction={decorate}] (8,0,-2)--(8,0,0);
        \end{scope}
        \path [fill, pattern=north east lines, pattern color = orange] plot (4,-4,0) .. controls (5,-3,0) and (7,3,0) .. (8,2,0)--plot (8,0,0)-- plot (8,0,0) .. controls (7,1,0)  and (5,-5,0)  .. (4,-6,0)--plot(4,-6,0);
        \fill[pattern=north east lines, pattern color = red](8,0,0)--(8,0,-2)--(8,2,-2)--(8,2,0);
        \node [red] at (-0.5,-1.5,0){$w^{n-\Hat{3}}(s_2)^{-1}$};
        \node [red!15!orange] at (-3.5,-1,0){$w^{n-\Hat{3}}(s_7)$};
        \node [red] at (8.8,0.8,-2){$B_{24}(n)$};
        \node [red!15!orange] at (7.5,-1.5,0){$B_{34}(n-\Hat{3})$};
        \node [blue] at (0,0.5,-2) [right]{$U(s_7,-4)$};
        \node at (-4,-6,0) [below]{$n-\Hat{3}$};
        \node at (0,0,0) [below]{$s_0=n$};
        \node at (2,0,0)[below]{$s_1$};
        \node at (2,0,-2)[below right]{$s_6$};
        \node at (0,0,-2)[below right]{$s_2$};
        \node at (2,2,0) [right] {$s_5$};
        \node at (2,2,-2)[right]{$s_4$};
        \node at (0,2,-2)[above]{$s_7$};
        \node at (0,1.8,0)[left]{$s_3$};
        \draw[->,>=stealth] (-4,0,0) -- (-3,0,0) node [right]{$x_1$};
        \draw[->,>=stealth] (-4,0,0) -- (-4,0,-1) node [above]{$x_2$};
        \draw[->,>=stealth] (-4,0,0) -- (-4,1,0) node [above]{$x_4$};
        \node at (-0.5,-7,0) {Construction of $u_{s_7s_2}^{n-\Hat{3}}$};
        \node at (7.5,-7,0) {Construction of $\Tilde{u}_{s_7s_2}^{n-\Hat{3}}$};
    \end{tikzpicture}
  \caption{Illustration for constructing the new components of the interpolation function. Here, we give an example of $u^{n-\hat{3}}_{s_7s_2}$. \textbf{(Left)} The original construction of $u^{n-\hat{3}}_{s_7s_2}$. The red line means $w^{n-\hat{3}}(s_7)$, and the blue line means the inverse of $w^{n-\hat{3}}(s_2)$. These functions are related to the interpolate function $S^{m}_{n,\mu}(x)$. \textbf{(Right)} The new construction of $\Tilde{u}^{n-\hat{3}}_{s_7s_2}$. We attach the $2$-form background gauge fields for corresponding plaquettes, for instance, $B_{24}(n)$ and $B_{34}(n-\hat{3})$, due to the $\mathbb{Z}_N$ $1$-form invariance.}
  \label{fig:redef_u}
\end{figure}

The fractionality of the topological charge arises from the $\mathbb{Z}_N$ $1$-form invariant action~\eqref{eq:Wilson_action} and the gauging procedure as\footnote{
    The $\mathbb{Z}_N$ $1$-form gauge transformation is as follows:
    \begin{align}
        U_\ell&\mapsto e^{\frac{2\pi i}{N}\lambda_\ell}U_\ell,
     \notag\\
        B_p&\mapsto B_p +(\dd{\lambda})_p\quad\bmod N,
     \label{eq:ZN_1form_gauge}
    \end{align}
    where $\lambda_\ell\in\mathbb{Z}_N$
    and~$(\dd{\lambda})_p=\sum_{\ell\in\partial p}\lambda_\ell$. Under this transformation, the action~\eqref{eq:Wilson_action_B} is invariant.
}
\begin{align}
    S_{\mathrm{W}}[U_\ell,B_p]
    =\sum_p\beta
    \left[\tr\left(\1-e^{-\frac{2\pi i}{N} B_p}U_p\right)
    +\mathrm{c.c.}\right],\quad \sum_{p\in\partial f}B_p=0\bmod N ,
    \label{eq:Wilson_action_B}
\end{align}
where we have defined the background $2$-form gauge field $B_p=B_{\mu\nu}(n)$ at the face of a plaquette. From Eq.~\eqref{eq:Wilson_action_B}, the $\mathbb{Z}_N$ $1$-form gauge invariance demands that each plaquette variable~$U_p$ should be multiplied by~$e^{-\frac{2\pi i}{N}B_p}$. Then, we reconstruct all components in the definition of the transition function $v_{n,\mu}(x)$ in the following: At first, we define the admissibility condition~\eqref{eq:admissibility} once more
\begin{align}
    \left\|\1-e^{-\frac{2\pi i}{N}B_p}U_p\right\|<\varepsilon,
    \label{eq:admissibilty_B}
\end{align}
where one finds that $\varepsilon\lesssim0.074$ is sufficiently small. Next, we redefine the interpolation function $S^m_{n,\mu}$ as $\Tilde{S}^m_{n,\mu}(x)$ in terms of $\Tilde{u}_{ss'}^m$ constructed by multiplication of appropriate $\mathbb{Z}_N$ plaquette fields~$e^{-\frac{2\pi i}{N}B_p}$; the illustration of this redefinition of the components in $S^m_{n,\mu}(x)$ is depicted in Fig.~\ref{fig:redef_u}. Finally, we see the gauge covariance of~$\Tilde{v}_{n,\mu}(x)$, that is, $\Tilde{v}_{n,\mu}(x)\mapsto e^{\frac{2\pi i}{N}\lambda_{\mu}(n-\Hat{\mu})}\Tilde{v}_{n,\mu}(x)$.

By using this gauge-covariant transition function $\Tilde{v}_{n,\mu}(x)$, we can check the cocycle condition,
\begin{align}
    \Tilde{v}_{n-\Hat{\nu},\mu}(x)\Tilde{v}_{n,\nu}(x)
    \Tilde{v}_{n,\mu}(x)^{-1}\Tilde{v}_{n-\Hat{\mu},\nu}(x)^{-1}
    =e^{\frac{2\pi i}{N}B_{\mu\nu}(n-\Hat{\mu}-\Hat{\nu})}\1.
\end{align}
This implies that under our construction of the transition function the principal bundle can possess a rich structure by an amount of~$\mathbb{Z}_N$ in the right hand side, where the topological charge is shifted by a fractional value in the unit of~$1/N$. Then, substituting $\Tilde{v}_{n,\mu}(x)$ instead of~$v_{n,\mu}(x)$ into~Eq.~\eqref{eq:TopologicalChargeDensity}, we find that the result is written in a local way by using the cohomological operations as
\begin{equation}
    Q_{\mathrm{top}}
    =-\frac{1}{N}\int_{T^4}\frac{1}{2}P_2(B_p)\qquad\bmod1
    \qquad\in\frac{1}{N}\mathbb{Z},
    \label{eq:fractionalcharge}
\end{equation}
where the Pontryagin square $P_2$ is defined in terms of (higher-)cup products by
\begin{equation}
    P_2(B_p)=B_p\cup B_p+B_p\cup_1\dd{B_p}.
\end{equation}

\subsection{'t~Hooft anomaly}
Using the above construction of the fractional topological charge on the lattice, let us consider the 't~Hooft anomaly in the lattice gauge theory. First, the definition of the partition function is given by
\begin{align}
    Z_{\theta}[B_p]
    =\int_{\mathfrak{A}_{\varepsilon}[B_p]}\dd{U_\ell}
    \exp\left(
    -S_{\mathrm{W}}[U_\ell,B_p]+i \theta Q_{\mathrm{top}}[U_\ell,B_p]
    \right),
    \label{eq:partition_function_B}
\end{align}
where we denote the set of admissible gauge fields as
\begin{align}
   \mathfrak{A}_{\varepsilon}[B_p]
   =\left\{
   \left\{U_\ell\right\}\,
   \Bigm|
   \text{$\left\|\1-e^{-\frac{2\pi i}{N}B_p}U_p\right\|<\varepsilon$ for
   all $p$}
   \right\}.
\end{align} 
Note that $\mathfrak{A}_{\varepsilon}[B_p]$ is $SU(N)$ gauge invariant and $\mathbb{Z}_N$ $1$-form gauge covariant.

Then, substituting the expression of $Q_{\mathrm{top}}$~\eqref{eq:fractionalcharge} into the partition function~\eqref{eq:partition_function_B} and carrying out the shift $\theta\to \theta+2\pi$, we obtain
\begin{align}
    Z_{\theta+2\pi}[B_p]
    =\exp\left[-\frac{2\pi i}{N}\int_{T^4}\frac{1}{2}P_2(B_p)\right]
    Z_{\theta}[B_p].
 \label{eq:lattice_tHooftAnomaly}
\end{align}
This violation of $2\pi$ periodicity of~$\theta$ reproduces the result of the 't~Hooft anomaly in the continuum theory between the shift symmetry of~$\theta\to\theta+2\pi$ and the $\mathbb{Z}_N$ $1$-form symmetry for $4$d $SU(N)$ pure Yang--Mills theory~\cite{Gaiotto:2017yup}.

\section{Conclusion and future works}
We construct the fractional topological charge in the lattice $SU(N)$ gauge theory coupled with the background $\mathbb{Z}_N$ 2-form gauge fields $B_p$ by extending L\"uscher's construction. Then, using this fractional topological charge on the lattice, we calculate the partition function of the lattice $SU(N)$ gauge theory with $B_p$ and conclude that this theory has the mixed 't~Hooft anomaly between the shift symmetry of~$\theta\to\theta+2\pi$ and the $\mathbb{Z}_N$ $1$-form symmetry.  This result implies that  the lattice gauge theory rigorously provides for the result of the mixed 't~Hooft anomaly between the $\mathbb{Z}_N$ $1$-form symmetry and the $\theta$ periodicity as a completely regularized theory.

In this talk, we only focus on $p$-form symmetries, but there are many studies of non-invertible symmetries which is the other aspects of generalized symmetries. For traditional symmetries, the symmetry generator possesses the inverse. As a generalization, we consider that the symmetry generator does not possess the inverse and we call it non-invertible symmetry. The non-invertible symmetry has the 't~Hooft anomaly as well as $p$-form symmetries~\cite{Cordova:2022ieu,Choi:2021kmx}. Then, our construction of the fractional topological charge would be useful to construct non-invertible symmetries on the lattice theory.

\subsection*{Acknowledgements}
This work was partially supported by JSPS KAKENHI Grant Numbers JP22KJ2096 (O.M.), JP20H01903 (H.S.), JP22H01218, and JP20K22350 (Y.T.). The work of Y.T. was supported by Center for Gravitational Physics and Quantum Information (CGPQI) at YITP. The work of M.A. was supported by Kyushu University Innovator Fellowship Program in Quantum Science Area.

\bibliographystyle{JHEP}
\bibliography{inspire}
\end{document}